\documentclass{article}
\usepackage{iclr2027_conference,times}

\usepackage{amsmath,amsfonts,bm}

\def\eqref#1{equation~\ref{#1}}

\def\1{\bm{1}}

\DeclareMathAlphabet{\mathsfit}{\encodingdefault}{\sfdefault}{m}{sl}
\SetMathAlphabet{\mathsfit}{bold}{\encodingdefault}{\sfdefault}{bx}{n}

\usepackage{graphicx}
\usepackage{booktabs,multirow,makecell,graphicx}
\usepackage{booktabs,tabularx,array}
\usepackage{needspace}
\usepackage{tikz}
\usetikzlibrary{arrows.meta,positioning,calc}
\usepackage{float}

\floatstyle{ruled}
\newfloat{algorithm}{tbp}{loa}
\floatname{algorithm}{Algorithm}

\usepackage{hyperref}
\usepackage{url}

\hypersetup{
    hidelinks,
    pdftitle={TeQCo: Temporal Quota Consistency for Training-Free Video Token Compression},
    pdfauthor={Yuqi Li, Xianglong Wang, Neil Rodrigues, Yiru Fang, Weilun Feng, Chuanyue Yan, Chuangguang Yang, Yingli Tian, Zhongwei Cheng, Wei Li, Lin Chen}
}

\providecommand{\tightlist}{
    \setlength{\itemsep}{0pt}
    \setlength{\parskip}{0pt}
}

\title{
TeQCo: Temporal Quota Consistency\\
for Training-Free Video Token Compression
}

\author{
Yuqi Li$^{1,2}$,
Xianglong Wang$^{1}$,
Neil Rodrigues$^{1}$,
Yiru Fang$^{1}$,
Weilun Feng$^{3}$,
Chuanyue Yan$^{4}$, \\
Chuangguang Yang$^{3}$,
Yingli Tian$^{2}$,
Zhongwei Cheng$^{1}$,
Wei Li$^{1}$,
Lin Chen$^{1}$ \\
\\
$^{1}$Wyze Labs, Inc. \\
$^{2}$The City College of New York, CUNY \\
$^{3}$Visual Information Processing and Learning \\
$^{4}$Wuhan Institute of Technology
}

\iclrfinalcopy

\begin{document}

\maketitle

\lhead{Preprint}

\begin{abstract}

Video token compression must control both temporal coverage and task relevance, but sequential stages can conflict: a content-driven allocation stage may be overwritten by a global ranking that removes entire temporal positions. We identify this as an allocation conflict and propose \textbf{TeQCo}, a training-free framework that resolves it through \textbf{temporal quota consistency}. TeQCo allocates candidate capacity from first-order changes in normalized frame representations and aggregates all tokens within each frame into a compact candidate set. Candidates are then scored by prior-corrected task attention and selected within integer frame quotas inherited from the first-stage counts. Content variation determines temporal capacity; question relevance determines which candidates fill it. This design guarantees an exact token budget and a per-frame coverage floor whenever feasible. On the 2,700-question Video-MME benchmark with Qwen3-VL-8B, TeQCo achieves 68.52\% accuracy at 510~ms TTFT, a 32.9\% TTFT reduction with a 1.00-percentage-point accuracy decrease.

\par\smallskip


\end{abstract}

\hypertarget{sec:introduction}{%
\section{Introduction}\label{sec:introduction}}

Video understanding requires integrating evidence about objects,
actions, and their changes across
time~\citep{nguyen2024videolanguagesurvey}.
Increasing the number of sampled frames or their spatial resolution
rapidly lengthens the visual token sequence and raises language-model
prefill costs~\citep{zhang2026efficientvlmsurvey}.
Video token compression must therefore control computation while
retaining the evidence needed to answer a
question~\citep{shao2026tokencompressionsurvey}. Prior work has developed training-free strategies based on frame-level budgets, spatiotemporal redundancy, and task-conditioned attention \citep{vcast,priortr,framefusion,vidcom2,fastvid,holitom,aim}.

These strategies address distinct decisions. Repeated backgrounds create temporal redundancy, whereas a brief change in object state may introduce useful evidence; within a frame, many regions may be irrelevant to the question. Compression must decide both \textbf{how much capacity each temporal position receives} and \textbf{which evidence occupies that capacity}. Uniform retention does not reflect changing video content, while global ranking by question attention does not explicitly protect context distributed across multiple moments.

We identify a structural conflict in sequential compression: \textbf{successive stages can make inconsistent decisions about the same temporal budget}. For example, suppose a content-driven stage retains \((6,4,2)\) tokens at three temporal positions and a subsequent stage reduces the total from 12 to 6. Unconstrained global Top-K selection can produce \((5,1,0)\), removing the third position entirely. Even informative relevance scores can therefore overwrite the allocation established upstream. The first stage assigns capacity across frames, while the second implicitly reallocates that capacity. Coordinating these decisions is essential to preserving a content-adaptive temporal distribution through successive reductions.

This distinction matters for relational questions. To determine whether an object changes state before or after an action, a model may need both the action and a comparatively quiet reference moment. A highly salient frame can provide strong local evidence while the reference establishes its temporal meaning. Treating temporal capacity as a shared constraint keeps such positions represented as the question refines their visual content.

We propose \textbf{TeQCo}, a training-free framework that connects content-adaptive compression and task-dependent evidence selection through \textbf{Temporal Quota Consistency}. TeQCo measures first-order changes in normalized frame representations to allocate candidate capacity, selects representative locations using local temporal differences, and aggregates similar features within each frame. It then evaluates candidates with prior-corrected task attention and selects within inherited frame quotas. This design assigns complementary roles to the two stages: content variation determines temporal capacity, and question relevance determines the evidence retained within each frame. Quota inheritance preserves this coordination through the final reduction.

Temporal quota consistency provides a concrete allocation rule: ideal final quotas are proportional to the actual upstream candidate counts, and bounded integer allocation enforces the exact final budget and frame capacities. A feasible coverage floor retains at least one token at every nonempty temporal position. Integer rounding and active coverage bounds translate the inherited proportions into realizable per-frame counts. In the example above, TeQCo assigns \((3,2,1)\) and selects the highest-scoring candidates within each frame, preserving all three temporal positions while halving the candidate sequence.

The contributions are threefold:

\begin{enumerate}
\def\labelenumi{\arabic{enumi}.}
\tightlist
\item
  \textbf{Cross-stage temporal quota consistency.} We identify that a task-driven global ranking stage can overwrite the temporal allocation established by upstream content-driven reduction, and resolve this by inheriting integer frame quotas from realized candidate counts, making temporal capacity a shared constraint across stages.
\item
  \textbf{Content-adaptive candidate construction.} We combine standardized first-order frame differences for temporal allocation, local temporal differences for representative selection, and within-frame aggregation, directing capacity toward changing content while ensuring every original token contributes to the candidate set.
\item
  \textbf{Exact-budget, quota-constrained evidence selection.} We derive a deterministic integer quota allocation and prove that per-frame top-K selection exactly maximizes the additive relevance objective under the assigned quotas (Appendix~\ref{app:optimality}), satisfying the final budget, frame capacities, and a feasible coverage floor with no additional learned components.
\end{enumerate}

Together, these components establish an explicit connection between temporal capacity and evidence identity. TeQCo preserves content-adaptive allocation across stages while using the question to refine which candidates carry that capacity into the final language-model prefill.

\section{Related Work}
\label{sec:related}

\textbf{Visual token reduction.} ToMe merges similar tokens through lightweight matching inside vision transformers \citep{tome}. LLaVA-PruMerge selects representatives using sparse visual attention and supplements them through similarity-based aggregation \citep{prumerge}. VisionZip selects dominant tokens using vision-encoder attention
and aggregates the remaining tokens into contextual representations~\citep{visionzip}. FastV instead uses early language-model attention to prune visual tokens from subsequent computation \citep{fastv}. SparseVLM uses visually relevant text tokens to guide progressive visual-token pruning, with layer-adaptive sparsification and token recycling~\citep{sparsevlm}. These approaches establish complementary compression signals: feature similarity supports early aggregation, while multimodal attention guides task-dependent selection. TeQCo connects these operations through a shared temporal allocation.

\textbf{Temporal redundancy and adaptive budgets.} FrameFusion exploits similarities between corresponding locations in adjacent frames, combining early merging with importance-based pruning \citep{framefusion}. FastVID uses temporal segmentation, density-based selection, and contextual merging to compress video structure \citep{fastvid}. VidCom$^2$ assigns different compression ratios to frames according to uniqueness and selects tokens using frame-level and video-level distinctiveness \citep{vidcom2}. V-CAST allocates frame budgets from semantic-trajectory curvature and retains spatial evidence at its original grid coordinates \citep{vcast}. These methods establish temporal organization and unequal frame capacity as useful components of video compression. TeQCo uses first-order representation changes to construct content-adaptive candidates and carries their realized frame proportions into question-dependent selection. The distinction from these methods is that TeQCo treats the
upstream temporal distribution as a binding constraint on the downstream selection: a subsequent global top-K can reduce a frame to zero tokens even when the upstream stage assigned it capacity proportional to its content change. The inherited quota mechanism prevents this by construction. 

\textbf{Task relevance and multi-stage coordination.} PriorTR contrasts task-conditioned attention with a causally isolated, question-free probe to obtain prior-corrected token scores \citep{priortr}. HoliTom combines spatiotemporal compression outside the language model with similarity-based merging inside it \citep{holitom}, while AIM couples pre-LLM merging with progressive importance-based pruning \citep{aim}. These approaches motivate combining content and task signals at different stages. TeQCo gives this combination an explicit allocation interface: the realized candidate counts from Stage 1 determine the per-frame integer quotas enforced in Stage 2. Frame-wise ranking selects the highest-scoring evidence within each assigned quota, so the content signal and task score govern distinct, non-competing decisions rather than implicitly competing over the same capacity.

\section{Method}
\label{sec:method}

\subsection{Overview}
\label{sec:overview}

TeQCo operates on the visual token sequence produced by a frozen vision-language model. Full visual encoding produces features $x_{t,i}\in\mathbb R^d$ for $T$ processed temporal positions and $P$ spatial tokens per position, giving $N=TP$ visual tokens in total. We use \emph{frame} to refer to a temporal position after temporal patch grouping and denote the number of nonvisual tokens by $M$. The compression target
is a final sequence of $K \ll N$ visual tokens for the language-model prefill, with all model parameters remaining frozen throughout.

\begin{figure}[t]
    \centering
    \includegraphics[width=\linewidth]{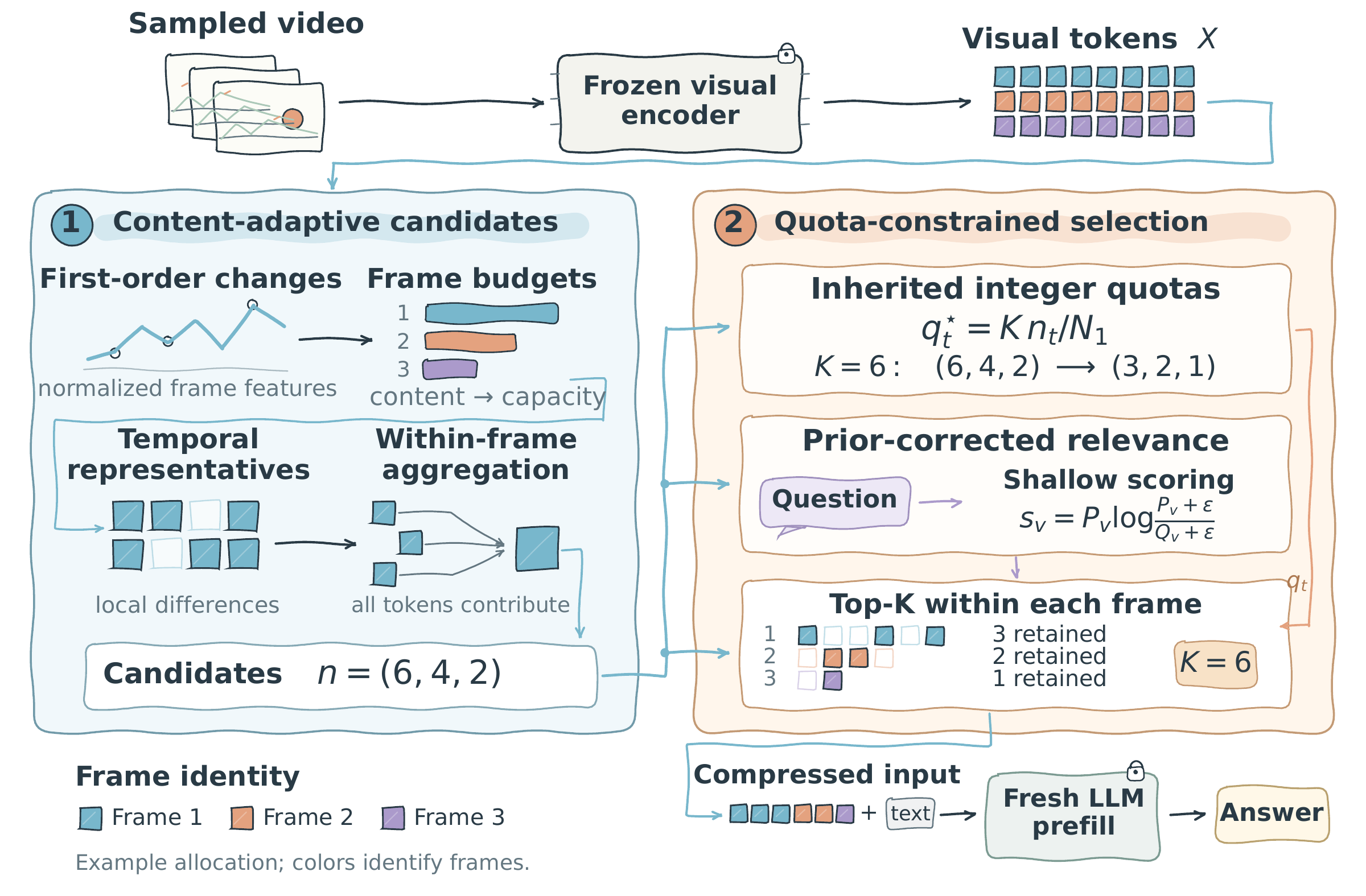}
    \caption{
\textbf{TeQCo pipeline with temporal quota consistency.}
\textbf{Stage 1} allocates candidate capacity using first-order changes in normalized frame representations, selects representatives through local temporal differences, and aggregates all tokens within each frame.
\textbf{Stage 2} inherits integer frame quotas from realized candidate counts.
A shallow pass compares task-conditioned attention with a question-free reference to compute prior-corrected relevance, selecting the top $q_t$ candidates per frame.
Selection enforces the exact budget and frame capacities, with per-frame coverage when feasible.
The example maps $n=(6,4,2)$ to $q=(3,2,1)$ for $K=6$.
Selected visual tokens and all nonvisual text enter a fresh language-model prefill for answer generation.
All model parameters remain frozen.
}
    \label{fig:pipeline}
\end{figure}

The first stage produces $n_t$ candidates per frame and $N_1=\sum_t n_t$ candidates overall. The second retains exactly $K$ visual tokens under inherited frame quotas. Figure~\ref{fig:pipeline} summarizes the flow from content-adaptive candidates to quota-constrained evidence selection.


\subsection{Content-Adaptive Candidate Construction}
\label{sec:allocation}\label{sec:aggregation}

\textbf{Temporal capacity.} We summarize each frame by its mean feature and measure first-order changes between normalized summaries:

\begin{equation}
\label{eq:frame-features}
\mu_t=\frac{1}{P}\sum_i x_{t,i},
\qquad
\delta_t=\|\hat\mu_t-\hat\mu_{t-1}\|_2,\quad t\geq2.
\end{equation}

A hat denotes L2 normalization. Displacement magnitude responds to frame-to-frame representation change, including sustained variation along a locally straight trajectory. Standardization puts these changes on a common scale for allocation:

\begin{equation}
\label{eq:frame-weights}
z_t=\frac{\delta_t-\bar\delta}{\max(\operatorname{std}(\delta),\epsilon)},
\qquad
p_t=\frac{\exp(z_t/\tau)}{\sum_j\exp(z_j/\tau)}.
\end{equation}

For nominal budget $B_1=\operatorname{round}(\alpha N)$, largest-remainder allocation converts $B_1p_t$ to integer counts $b_t$, followed by $n_t=\max\{1,\min(P,b_t)\}$. The realized $n_t$ and $N_1$ capture the effect of these bounds and define the capacity inherited by the second stage. Boundary handling is specified in Appendix~\ref{app:normalization}.

Normalization makes allocation respond to feature direction rather than magnitude. Standardization then expresses each displacement relative to the variation within its video, while $\tau$ controls how strongly this variation changes capacity. Equal displacements yield uniform weights; increasing $\tau$ moves the allocation toward this balanced regime. Thus, adaptation redistributes a fixed nominal budget rather than increasing computation when visual change grows.

\textbf{Representatives and aggregation.} We rank spatial locations by their change from the corresponding location in the preceding frame:

\begin{equation}
\label{eq:local-differences}
c_{t,i}=1-\langle\hat x_{t,i},\hat x_{t-1,i}\rangle,\qquad t\geq2.
\end{equation}

The $n_t$ highest-scoring locations form the representative set $\mathcal R_t$. Each remaining token is assigned to its most similar representative in the same frame. For the resulting cluster $\mathcal C_{t,j}$, the candidate is

\begin{equation}
\label{eq:aggregation}
y_{t,j}=\frac{1}{|\mathcal C_{t,j}|}\sum_{i\in\mathcal C_{t,j}}x_{t,i}.
\end{equation}

Temporal differences prioritize changing regions, while aggregation incorporates the remaining visual context. Every original token contributes to one candidate cluster. For backbones with auxiliary visual streams such as DeepStack, shared aggregation maps align these streams with the main embeddings; representative grid indices provide their spatial positions (Appendix~\ref{app:alignment}). The two change signals operate at different resolutions: pooled displacement determines how many candidates a frame receives, whereas local differences determine their representative locations. Averaging within each resulting cluster allows relatively static appearance to contribute alongside the selected changing regions.

\subsection{Quota-Constrained Task Selection}
\label{sec:scoring}\label{sec:quotas}\label{sec:selection}

\textbf{Task relevance.} Following PriorTR \citep{priortr}, we insert a question-free probe after the visual sequence and before the text suffix. A shallow forward pass estimates reference attention $Q_v$ from the probe and task-conditioned attention $P_v$ from the subsequent text, both normalized over visual candidates. The candidate score is

\begin{equation}
\label{eq:relevance}
s_v=P_v\log\frac{P_v+\epsilon}{Q_v+\epsilon}.
\end{equation}

Causal masking makes $Q_v$ depend on the preceding video and prompt, while $P_v$ incorporates the question. This signed ranking statistic combines task attention with its relative emphasis over the reference. Appendix~\ref{app:attention} specifies the attention aggregation.

\textbf{Inherited quotas.} For second-stage retention ratio $\beta$, we set the final budget and ideal frame quotas to

\begin{equation}
\label{eq:final-budget}
\begin{gathered}
K=\min\{N_1,\max(1,\operatorname{round}(\beta N_1))\},  
\qquad
q_t^*=K\frac{n_t}{N_1}.
\end{gathered}
\end{equation}

These quotas carry the realized candidate proportions into task-dependent selection. With coverage floor $\ell_t=\min(m,n_t)$ and sufficient budget $K\geq\sum_t\ell_t$, bounded integer allocation gives

\begin{equation}
\label{eq:quota-constraints}
\sum_t q_t=K,
\qquad \ell_t\leq q_t\leq n_t.
\end{equation}

We initialize from bounded, rounded-down ideal quotas and correct the total by allocating or reclaiming slots according to the largest deficit or surplus. When the coverage floor is infeasible, setting $\ell_t=0$ preserves exact budget control. Appendix~\ref{app:quotas} gives the complete deterministic procedure.

\textbf{Evidence selection.} Let $\mathcal Y_t$ denote the candidates in frame $t$. We retain

\begin{equation}
\label{eq:selection}
\mathcal S_t=\operatorname{TopK}_{v\in\mathcal Y_t}(s_v,q_t).
\end{equation}

For fixed scores and quotas, this selection exactly maximizes the additive relevance objective (Appendix~\ref{app:optimality}). It satisfies the final budget and frame capacities, with per-frame coverage whenever feasible. Quotas fix temporal capacity, while the question determines candidate identity. The retained indices jointly select visual embeddings, DeepStack features, and positions; all nonvisual text is preserved and the probe is removed.

\textbf{Local ranking suffices.} For fixed candidates and quotas, selection is unchanged by any frame-specific transformation $s'_v=a_t s_v+b_t$ with $a_t>0$ and $v\in\mathcal Y_t$. Such transformations preserve each frame's ordering and therefore its selected candidates (Appendix~\ref{app:score-invariance}). TeQCo consequently needs relevance scores to distinguish evidence within frames, rather than to calibrate competition between frames. A frame-wide score offset cannot consume another frame's allocation. This property makes the division between content-adaptive capacity and question-dependent evidence explicit in the selection operator itself.

\subsection{Execution and Computational Benefit}
\label{sec:execution}

TeQCo separates shallow relevance estimation from full-depth inference. The scoring pass evaluates the first three decoder blocks and reads scores at the third block's input. A fresh prefill then processes the selected input from the first layer, concentrating full-depth computation on $K+M$ tokens instead of $N+M$.

The default $\alpha=\beta=0.5$ targets $K/N\approx25\%$ visual retention; $\tau=8$ controls content adaptation and $m=1$ provides feasible per-frame coverage. Cost accounting includes full visual encoding, candidate construction, shallow scoring, quota selection, and final prefill. Appendix~\ref{app:procedure-cost} provides the algorithm and cost decomposition; Appendix~\ref{app:configuration} lists the configuration and an allocation-deviation diagnostic.

\section{Experiments}
\label{sec:experiments}
\subsection{Experimental Setup}

\paragraph{Benchmarks and Models.}
We evaluate on Video-MME~\citep{videomme}, EgoSchema~\citep{egoschema}, LongVideoBench~\citep{longvideobench}, and MVBench~\citep{mvbench}, using Qwen3-VL-8B and Qwen3-VL-2B~\citep{qwen3vl}, InternVL3.5-8B~\citep{internvl35}, and LLaVA-OneVision-7B~\citep{llavaonevision}. Video-MME contains 2,700 questions, divided equally among short, medium, and long videos. We use the 500 labelled EgoSchema questions, 1,337 LongVideoBench validation questions, and 3,800 MVBench questions.

\paragraph{Research Questions.}
We assess the accuracy--latency tradeoff across models and tasks (RQ1), its relationship to total latency and memory (RQ2), the interaction between candidate construction and final selection (RQ3), and sensitivity to video duration (RQ4).

\paragraph{Input Sampling.}
Both Qwen models use 1 FPS sampling with 4--768 frames and a maximum spatial edge of 512 pixels. InternVL and LLaVA use 32 frames per video, with spatial settings of 448 and 384, respectively. Tables~\ref{tab:mergedresults} and~\ref{tab:fixedframesresults} report these protocols separately.

\paragraph{Compression Settings.}
All full-pipeline TeQCo results use inherited integer frame quotas and frame-wise selection as specified in Section~\ref{sec:selection}. The Qwen configuration uses $\alpha=\beta=0.5$ and $\tau=8$; InternVL and LLaVA use $\tau=3$. The per-frame coverage floor is $m=1$ whenever feasible. Comparisons use each method's reported operating point, jointly considering task accuracy and measured latency.

\paragraph{Implementation and Metrics.}
All runs use NVIDIA H100 80GB HBM3 GPUs, batch size 1, and warmup. We report accuracy in percent and latency in milliseconds. TTFT includes visual encoding, compression, shallow scoring, selection, and final prefill; video decoding and preprocessing precede this interval. End-to-end latency (E2E) also includes preprocessing and answer generation. Memory is peak allocated device memory over the execution. Each latency comparison uses the corresponding full-input baseline.

\begin{table}[t]
\centering
\caption{\textbf{Qwen3-VL results with input sampling at 1 FPS.} Higher accuracy (\%) and lower TTFT (ms) are better.}
\label{tab:mergedresults}
\label{tab:expbaselines}
\label{tab:expbaselines2b}
\begingroup
\fontsize{8}{8.5}\selectfont
\setlength{\tabcolsep}{0.6pt}
\setlength{\aboverulesep}{0.7pt}
\setlength{\belowrulesep}{0.7pt}
\renewcommand{\arraystretch}{1.0}
\begin{tabular*}{\linewidth}{@{\extracolsep{\fill}}l*{14}{c}@{}}
\toprule
 & \multicolumn{8}{c}{\textbf{Video-MME}} & \multicolumn{2}{c}{\textbf{EgoSchema}} & \multicolumn{2}{c}{\textbf{LongVideo}} & \multicolumn{2}{c}{\textbf{MVBench}} \\
\cmidrule(lr){2-9}
 & \multicolumn{2}{c}{Short} & \multicolumn{2}{c}{Medium} & \multicolumn{2}{c}{Long} & \multicolumn{2}{c}{All} & \multicolumn{2}{c}{} & \multicolumn{2}{c}{\textbf{Bench}} & \multicolumn{2}{c}{(RGB)} \\
\cmidrule(lr){2-3}\cmidrule(lr){4-5}\cmidrule(lr){6-7}\cmidrule(lr){8-9}\cmidrule(lr){10-11}\cmidrule(lr){12-13}\cmidrule(lr){14-15}
\textbf{Method} & \makebox[2.3em]{Acc.$\uparrow$} & \makebox[3.1em]{TTFT$\downarrow$} & \makebox[2.3em]{Acc.$\uparrow$} & \makebox[3.1em]{TTFT$\downarrow$} & \makebox[2.3em]{Acc.$\uparrow$} & \makebox[3.1em]{TTFT$\downarrow$} & \makebox[2.3em]{Acc.$\uparrow$} & \makebox[3.1em]{TTFT$\downarrow$} & \makebox[2.3em]{Acc.$\uparrow$} & \makebox[3.1em]{TTFT$\downarrow$} & \makebox[2.3em]{Acc.$\uparrow$} & \makebox[3.1em]{TTFT$\downarrow$} & \makebox[2.3em]{Acc.$\uparrow$} & \makebox[3.1em]{TTFT$\downarrow$} \\
\midrule
\multicolumn{15}{c}{\textbf{Qwen3-VL-8B}} \\
\midrule
Full input & 79.00 & 342 & 69.22 & 877 & 60.33 & 1062 & 69.52 & 760 & 72.20 & 716 & 62.08 & 873 & 66.58 & 86 \\
Curv. merge & \textbf{78.56} & 244 & \underline{68.00} & 687 & 58.89 & 865 & 68.48 & 598 & 69.60 & 489 & \textbf{61.27} & 698 & \textbf{64.79} & 70 \\
PriorTR & 78.33 & 262 & \textbf{68.67} & 713 & 58.89 & 899 & \textbf{68.63} & 625 & \textbf{70.80} & 546 & \underline{60.86} & 730 & \underline{64.56} & 78 \\
EarlyTom & 76.11 & \textbf{183} & 66.56 & \textbf{473} & \textbf{59.22} & \textbf{597} & 67.30 & \textbf{418} & 69.20 & \textbf{369} & 58.69 & 641 & 62.11 & \textbf{58} \\
FrameFusion & 75.89 & 355 & 67.11 & 815 & 58.22 & 967 & 67.07 & 712 & 69.80 & 699 & 60.14 & 821 & 63.03 & 149 \\
VidCom$^{2}$ & 74.33 & 197 & 65.67 & 581 & 56.56 & 772 & 65.52 & 517 & 68.40 & \underline{387} & 58.07 & \textbf{457} & 60.26 & 68 \\
FastVID & 73.56 & 217 & 65.00 & 681 & 55.11 & 935 & 64.56 & 611 & 69.40 & 419 & 58.42 & \underline{513} & 63.21 & 75 \\
HoliTom & 73.11 & 257 & 62.56 & 1746 & 55.00 & 3047 & 63.56 & 1683 & 70.00 & 554 & 57.25 & 656 & 61.18 & 74 \\
AIM & 73.56 & 386 & 63.00 & 1828 & 56.56 & 2917 & 64.37 & 1711 & 69.00 & 1063 & 59.13 & 1204 & 60.45 & 103 \\
\addlinespace[1pt]
\textbf{TeQCo} & \underline{78.44} & \underline{191} & \underline{68.00} & \underline{575} & \underline{59.11} & \underline{764} & \underline{68.52} & \underline{510} & \underline{70.40} & 421 & 58.19 & 688 & 62.11 & \underline{67} \\
\midrule
\multicolumn{15}{c}{\textbf{Qwen3-VL-2B}} \\
\midrule
Full input & 73.89 & 175 & 60.44 & 489 & 48.00 & 609 & 60.78 & 424 & 64.20 & 351 & 54.45 & 528 & 60.84 & 57 \\
Curv. merge & 72.22 & 139 & \underline{59.44} & 421 & \underline{47.89} & 570 & 59.85 & 376 & \textbf{63.60} & 279 & 53.56 & 487 & \underline{59.26} & \underline{50} \\
PriorTR & \underline{73.56} & 149 & \textbf{60.11} & 428 & 47.67 & 555 & \textbf{60.44} & 377 & \underline{63.20} & 296 & 53.89 & 504 & \textbf{61.58} & 52 \\
EarlyTom & 69.67 & \textbf{95} & 56.56 & \textbf{259} & 47.56 & \textbf{338} & 57.93 & \textbf{231} & 59.00 & \textbf{184} & 52.28 & 269 & 57.53 & \textbf{40} \\
FrameFusion & 71.11 & 279 & 58.33 & 536 & 47.78 & 669 & 59.07 & 495 & 62.20 & 446 & 54.47 & 559 & 57.26 & 126 \\
VidCom$^{2}$ & 71.22 & 133 & 59.22 & 405 & \textbf{48.00} & 532 & 59.48 & 357 & 62.40 & 247 & 53.68 & \textbf{298} & 56.76 & 55 \\
FastVID & 70.78 & 148 & 57.56 & 460 & 46.22 & 635 & 58.19 & 414 & 62.00 & 271 & \underline{56.26} & \underline{354} & 57.13 & 62 \\
HoliTom & 67.33 & 206 & 56.22 & 1594 & 46.67 & 2975 & 56.74 & 1592 & 61.80 & 457 & 55.12 & 583 & 56.84 & 65 \\
AIM & 67.22 & 217 & 57.11 & 899 & 46.89 & 1374 & 57.07 & 830 & 61.80 & 516 & \textbf{56.43} & 622 & 57.21 & 79 \\
\addlinespace[1pt]
\textbf{TeQCo} & \textbf{73.67} & \underline{127} & 58.56 & \underline{398} & \underline{47.89} & \underline{523} & \underline{60.04} & \underline{349} & 62.40 & \underline{242} & 53.40 & 487 & 58.00 & 53 \\
\bottomrule
\end{tabular*}
\par\vspace{3pt}
\begin{minipage}{\linewidth}
\fontsize{8}{8.5}\selectfont
\textit{Notes.} \textbf{Bold} and \underline{underlined} values indicate the best and second-best results in each column within each backbone, excluding Full input; ties share the same marking.

\end{minipage}
\endgroup
\end{table}

\begin{table}[t]
\centering
\caption{\textbf{InternVL3.5 and LLaVA-OneVision results with input sampling fixed at 32 frames.} Higher accuracy (\%) and lower TTFT (ms) are better.}
\label{tab:fixedframesresults}
\label{tab:expmodels}
\begingroup
\fontsize{8}{8.5}\selectfont
\setlength{\tabcolsep}{0.6pt}
\setlength{\aboverulesep}{0.7pt}
\setlength{\belowrulesep}{0.7pt}
\renewcommand{\arraystretch}{1.0}
\begin{tabular*}{\linewidth}{@{\extracolsep{\fill}}l*{14}{c}@{}}
\toprule
 & \multicolumn{8}{c}{\textbf{Video-MME}} & \multicolumn{2}{c}{\textbf{EgoSchema}} & \multicolumn{2}{c}{\textbf{LongVideo}} & \multicolumn{2}{c}{\textbf{MVBench}} \\
\cmidrule(lr){2-9}
 & \multicolumn{2}{c}{Short} & \multicolumn{2}{c}{Medium} & \multicolumn{2}{c}{Long} & \multicolumn{2}{c}{All} & \multicolumn{2}{c}{} & \multicolumn{2}{c}{\textbf{Bench}} & \multicolumn{2}{c}{(RGB)} \\
\cmidrule(lr){2-3}\cmidrule(lr){4-5}\cmidrule(lr){6-7}\cmidrule(lr){8-9}\cmidrule(lr){10-11}\cmidrule(lr){12-13}\cmidrule(lr){14-15}
\textbf{Method} & \makebox[2.3em]{Acc.$\uparrow$} & \makebox[3.1em]{TTFT$\downarrow$} & \makebox[2.3em]{Acc.$\uparrow$} & \makebox[3.1em]{TTFT$\downarrow$} & \makebox[2.3em]{Acc.$\uparrow$} & \makebox[3.1em]{TTFT$\downarrow$} & \makebox[2.3em]{Acc.$\uparrow$} & \makebox[3.1em]{TTFT$\downarrow$} & \makebox[2.3em]{Acc.$\uparrow$} & \makebox[3.1em]{TTFT$\downarrow$} & \makebox[2.3em]{Acc.$\uparrow$} & \makebox[3.1em]{TTFT$\downarrow$} & \makebox[2.3em]{Acc.$\uparrow$} & \makebox[3.1em]{TTFT$\downarrow$} \\
\midrule
\multicolumn{15}{c}{\textbf{InternVL3.5-8B}} \\
\midrule
Full input & 77.67 & 367 & 62.56 & 367 & 51.67 & 367 & 63.96 & 367 & 69.80 & 369 & 60.58 & 369 & 73.34 & 365 \\
Curv. merge & 75.78 & 246 & 62.22 & 249 & 52.11 & 251 & 63.37 & 249 & \textbf{70.40} & 243 & \underline{60.47} & 248 & 66.84 & 235 \\
PriorTR & \textbf{77.67} & 291 & \textbf{63.11} & 296 & 51.11 & 298 & \textbf{63.96} & 295 & \underline{69.80} & 302 & \textbf{60.63} & 298 & \underline{69.68} & 288 \\
EarlyTom & 74.22 & 188 & 59.78 & 188 & \underline{52.22} & 189 & 62.07 & 188 & 65.00 & 193 & 59.39 & 197 & 68.97 & 190 \\
FrameFusion & 75.89 & 358 & 60.33 & 302 & 51.78 & 284 & 62.67 & 315 & 68.60 & 367 & 59.88 & 364 & 68.82 & 311 \\
VidCom$^{2}$ & 74.33 & \textbf{149} & 60.56 & \textbf{149} & 51.11 & \textbf{149} & 62.00 & \textbf{149} & 67.80 & \textbf{153} & 58.74 & \textbf{149} & \textbf{70.61} & \textbf{149} \\
FastVID & 72.00 & \underline{175} & 58.33 & 187 & 49.89 & 191 & 60.07 & \underline{184} & 68.20 & \underline{173} & 55.32 & 192 & 67.71 & \underline{150} \\
HoliTom & 68.89 & 203 & 56.44 & \underline{186} & 48.44 & \underline{183} & 57.93 & 191 & 66.20 & 190 & 53.46 & 193 & 63.18 & 215 \\
AIM & 71.22 & 422 & 60.00 & 423 & 51.00 & 426 & 60.74 & 423 & 68.60 & 448 & 55.92 & 440 & 64.66 & 417 \\
\addlinespace[1pt]
\textbf{TeQCo} & \underline{76.67} & 182 & \underline{62.33} & 187 & \textbf{52.33} & 185 & \underline{63.77} & 185 & 68.40 & 193 & 58.94 & \underline{187} & 66.13 & 186 \\
\midrule
\multicolumn{15}{c}{\textbf{LLaVA-OneVision-7B}} \\
\midrule
Full input & 71.11 & 293 & 56.00 & 295 & 48.33 & 295 & 58.48 & 294 & 64.20 & 298 & 55.20 & 301 & 58.39 & 294 \\
Curv. merge & 70.11 & 232 & \textbf{56.56} & 235 & 47.00 & 239 & \textbf{57.89} & 236 & \textbf{64.60} & 239 & \underline{54.76} & 240 & \underline{58.03} & 236 \\
PriorTR & 69.78 & 291 & \underline{56.00} & 293 & 47.33 & 298 & 57.70 & 294 & 63.00 & 296 & 54.27 & 291 & 57.89 & 255 \\
EarlyTom & 69.44 & \textbf{157} & 55.44 & \textbf{160} & \textbf{48.67} & \textbf{161} & \underline{57.85} & \textbf{160} & 62.80 & \textbf{163} & 54.00 & \textbf{160} & 56.08 & \textbf{161} \\
FrameFusion & \textbf{70.67} & 350 & 55.44 & 341 & 47.44 & 335 & \underline{57.85} & 342 & 63.40 & 386 & \textbf{54.93} & 337 & \textbf{58.39} & 275 \\
VidCom$^{2}$ & 69.00 & \underline{169} & 55.56 & \underline{169} & \underline{47.89} & \underline{169} & 57.48 & \underline{169} & \underline{64.20} & \underline{173} & 53.84 & \underline{169} & 56.29 & \underline{169} \\
FastVID & 68.44 & 183 & 55.33 & 189 & 46.56 & 190 & 56.78 & 187 & \textbf{64.60} & 183 & 53.20 & 189 & 56.50 & \underline{169} \\
HoliTom & 68.00 & 225 & 54.78 & 210 & 47.33 & 208 & 56.70 & 214 & 64.00 & 219 & 52.68 & 220 & 57.47 & 234 \\
AIM & 67.78 & 256 & 54.78 & 257 & 47.78 & 260 & 56.78 & 258 & 63.20 & 272 & 52.32 & 271 & 56.24 & 255 \\
\addlinespace[1pt]
\textbf{TeQCo} & \underline{70.33} & 200 & 54.67 & 202 & \underline{47.89} & 201 & 57.63 & 201 & 62.40 & 202 & 53.18 & 200 & 56.68 & 199 \\
\bottomrule
\end{tabular*}
\par\vspace{3pt}
\begin{minipage}{\linewidth}
\fontsize{8}{8.5}\selectfont
\textit{Notes.} \textbf{Bold} and \underline{underlined} values indicate the best and second-best results in each column within each backbone, excluding Full input; ties share the same marking.
-- denotes an unavailable or unverified result.
Input sampling uses 32 frames for both backbones.
Methods are compared at their reported operating points.
All covers 2,700 Video-MME questions.
\end{minipage}
\endgroup
\end{table}

\subsection{RQ1: Task Performance and Latency}
\label{sec:rq1}

\paragraph{A compact sequence with useful task evidence.}
Tables~\ref{tab:mergedresults} and~\ref{tab:fixedframesresults} show lower TTFT with limited changes in full Video-MME accuracy across all four backbones. On both Qwen models, TeQCo improves accuracy and latency jointly over several compression baselines. Relative to PriorTR, it trades a small accuracy difference for faster first-token delivery. This comparison is consistent with the benefit of reducing the sequence before task scoring: relevance estimation operates on aggregated candidates, and full-depth inference processes a further quota-constrained subset. The result is a useful balance between retaining task evidence and reducing the work needed to access it.

\paragraph{Complementary capacity and relevance signals.}
TeQCo gives each sampled temporal position a feasible share of the final sequence, then uses the question to choose the evidence within that share. This structure is well suited to retaining a mixture of prominent events and quieter visual context. The strong Video-MME tradeoff on InternVL further shows that the same allocation principle transfers beyond the Qwen backbone. LLaVA exhibits a closer competition among operating points, with some methods offering faster inference or slightly higher accuracy. Across these backbones, the same quota rule yields compact sequences, while its benefit depends on the decoder and visual representation.

\paragraph{Task-dependent evidence density.}
The larger Qwen3-VL-8B accuracy gaps on LongVideoBench and MVBench, compared with Video-MME, show that accuracy retention varies by task. A mechanism consistent with this pattern is variation in the detail required from each retained frame. A retained frame supplies a temporal anchor, while its candidates determine how much within-frame evidence remains accessible. This distinction offers a mechanism-level interpretation of the task variation: preserving temporal coverage is most effective when compact candidates retain the discriminative content needed by the question. The tables identify where this compact representation delivers the strongest practical tradeoff.

\subsection{RQ2: Inference Latency and Memory}
\label{sec:rq2}

Table~\ref{tab:expcost} compares TeQCo with full-input inference using identical question IDs and reference answers. TTFT improves in every listed configuration, including the cost of constructing and scoring candidates. The consistently larger relative gains for Qwen3-VL-8B than for Qwen3-VL-2B fit the computational decomposition in Appendix~\ref{app:cost}: reducing full-depth decoder work has greater leverage when that work accounts for more of the inference path. The shallow pass is an additional investment that enables a smaller sequence throughout the final prefill.

\paragraph{First-token latency and total response time.}
Both models improve E2E latency on EgoSchema and LongVideoBench, whereas Video-MME has a small E2E increase despite faster TTFT. These metrics cover different portions of the execution. In particular, the much larger E2E times on Video-MME and LongVideoBench show that substantial runtime lies outside the first-token interval. Improvements to compressed prefill therefore contribute differently to the total response time across workloads. The Video-MME increase means that runtime outside this interval outweighs the measured TTFT saving; candidate and scoring costs are already included in TTFT. This separates the consistent improvement in first-token delivery from the workload-dependent change in total response time.

\paragraph{Peak memory reflects the complete execution.}
Peak memory decreases for the 8B model and increases slightly for the 2B model. Compression shortens the final sequence while leaving model weights unchanged. Moreover, the peak is determined by the largest live allocation across visual encoding, candidate scoring, and generation. The scoring implementation constructs a dense causal mask and temporarily retains candidate features (Appendix~\ref{app:cost}), so a smaller final sequence need not produce a proportionate change in the overall peak. The measurements thus establish latency reduction as the clearest benefit of the current execution path.

\begin{table}[t]
\centering\small
\setlength{\tabcolsep}{4pt}
\caption{Inference latency and peak memory for quota-constrained TeQCo and full-input inference on six Qwen configurations. B and T denote the full-input baseline and TeQCo, respectively. Latency is in ms; memory is peak allocated device memory in GB. Each pair uses identical question IDs and reference answers.}
\label{tab:expcost}
\sbox0{%
\begin{tabular}{llllllll}
\toprule
 & & \multicolumn{2}{c}{TTFT (ms)} & \multicolumn{2}{c}{E2E (ms)} & \multicolumn{2}{c}{Memory (GB)} \\
\cmidrule(lr){3-4}\cmidrule(lr){5-6}\cmidrule(lr){7-8}
Model & Benchmark & B & T & B & T & B & T \\
\midrule
8B & Video-MME & 760 & 510 & 11996.8 & 12112.8 & 22.61 & 21.69 \\
8B & EgoSchema & 716 & 421 & 2829.9 & 2273.7 & 21.80 & 21.14 \\
8B & LongVideoBench & 873 & 688 & 11803.4 & 11615.8 & 23.33 & 23.28 \\
2B & Video-MME & 424 & 349 & 11692.2 & 11910.8 & 7.68 & 7.72 \\
2B & EgoSchema & 351 & 242 & 2308.8 & 2106.9 & 7.13 & 7.29 \\
2B & LongVideoBench & 528 & 487 & 11504.5 & 11460.6 & 8.15 & 8.33 \\
\bottomrule
\end{tabular}}
\ifdim\wd0>\linewidth\resizebox{\linewidth}{!}{\usebox0}\else\usebox0\fi
\end{table}

\subsection{RQ3: Contributions of Candidate Construction and Final Selection}
\label{sec:rq3}

Table~\ref{tab:expablation} examines allocation rules and their composition with task-guided selection. Candidate-only rows retain the first-stage sequence; combined rows further reduce it using inherited quotas. These comparisons evaluate the transition from content-based aggregation to the complete, smaller-budget inference path.

\paragraph{Allocation determines the candidate evidence pool.}
All candidate-construction variants lower TTFT relative to full input. Their accuracy ordering changes with duration: curvature performs better on short and medium videos, while first-order allocation performs better on long videos. The two signals emphasize different aspects of a representation trajectory. Curvature highlights changes in direction, whereas displacement remains responsive to sustained content variation along a locally straight path. First-order allocation therefore supplies a complementary basis for distributing capacity across an extended sequence, with local representative selection refining the evidence within each frame.

\paragraph{A second reduction can retain the more useful evidence.}
Compared with first-order candidates alone, the full TeQCo pipeline is faster on every duration subset, improves accuracy on short and medium videos, and gives nearly unchanged accuracy on long videos. The transition shows that retaining more aggregated candidates is not always necessary for stronger task performance. Candidate construction organizes visual content; prior-corrected ranking then chooses which of those summaries to expose to the final decoder. Quota inheritance coordinates this refinement with the upstream temporal allocation. The resulting pattern supports evaluating the two stages as a composed inference strategy: they address redundancy and task relevance at different points in the computation.

\begin{table}[t]
\centering\small
\setlength{\tabcolsep}{4pt}
\caption{Qwen3-VL-8B component experiments with 900 questions per duration subset. Variants with final selection enforce inherited frame quotas. The Full input and TeQCo rows reproduce Table~\ref{tab:mergedresults}; the other rows report component measurements. Rows without scoring retain the candidate sequence, so adding quota-constrained selection also changes the token budget.}
\label{tab:expablation}
\sbox0{%
\begin{tabular}{lllllll}
\toprule
Variant & Short acc.(\%) & TTFT (ms) & Medium acc. (\%) & TTFT (ms) & Long acc. (\%) & TTFT (ms)\\
\midrule
Full input & 79.00 & 342 & 69.22 & 877 & 60.33 & 1062 \\
Fixed allocation & 77.89 & 243.3 & 67.67 & 673.5 & 58.78 & 844.2 \\
Curvature & 78.56 & 244.1 & 68.00 & 687.3 & 58.89 & 865.2 \\
Curvature + scoring & 77.89 & 210.6 & 67.00 & 619.5 & 57.33 & 800.9 \\
First order & 77.78 & 243.0 & 67.44 & 675.5 & 59.22 & 857.2 \\
TeQCo (full pipeline) & 78.44 & 191 & 68.00 & 575 & 59.11 & 764 \\
\bottomrule
\end{tabular}}
\ifdim\wd0>\linewidth\resizebox{\linewidth}{!}{\usebox0}\else\usebox0\fi
\end{table}

\subsection{RQ4: Sensitivity to Video Duration}
\label{sec:rq4}

With the sampling protocol fixed within each backbone, both Qwen models retain TTFT gains across short, medium, and long videos, but relative savings decrease with duration. Longer inputs increase both the work removed from the final prefill and the work retained in visual encoding, candidate construction, and shallow scoring. Consequently, the fraction of the full execution that compression eliminates, rather than video duration alone, determines the relative speedup. TeQCo's duration trend is consistent with these competing contributions to the same timed inference path.

Under fixed 32-frame sampling, InternVL and LLaVA show closely grouped TTFT values across duration subsets. Here, increasing source-video duration changes the temporal spacing of observations without proportionally increasing the sampled frame count. This provides a useful operational contrast with duration-dependent Qwen sampling: semantic video length and decoder input length are distinct workload properties. We interpret each trend within its backbone and preprocessing configuration.

The coverage condition also links temporal density to the final budget: with $m=1$, retaining every processed frame requires $K\geq T$. Once feasible, inherited quotas use the remaining capacity to refine frame representations according to content variation. This makes coverage an explicit constraint rather than an incidental outcome of ranking.

Accuracy changes are also non-monotonic with duration, particularly for Qwen3-VL-2B. More elapsed video time does not directly specify how much evidence the question requires or how that evidence is distributed.

\section{Conclusion}

TeQCo makes temporal capacity a shared constraint across sequential video-token compression stages. Content-adaptive aggregation defines a compact candidate pool, and inherited integer quotas guide question-dependent selection within each frame. This separation combines exact budget control, feasible temporal coverage, and optimal frame-wise selection for the fixed-score objective, using frozen model parameters. Experiments across four backbones show that the resulting compact sequences enable faster first-token delivery while retaining useful task evidence. Together, the results illustrate how successive stages can specialize in different signals while retaining a shared temporal allocation.

\section*{AI Use Statement}
We used ChatGPT (OpenAI) to assist with manuscript organization,
language polishing, and LaTeX formatting. It also assisted with
interpreting the author-provided experimental results and formulating
the frame-wise score-calibration property and its proof.
All reported experimental measurements were supplied by the authors.
The authors take full responsibility for the final manuscript,
including its scientific claims, mathematical arguments, and
experimental results.

\bibliography{iclr2027_conference}
\bibliographystyle{iclr2027_conference}

\clearpage
\appendix

\numberwithin{equation}{section}
\renewcommand{\theequation}{\thesection\arabic{equation}}
\renewcommand{\theHequation}{appendix.\thesection.\arabic{equation}}

\section{Algorithmic Details and Selection Properties}
\label{app:algorithmic-details}

\subsection{Normalization, Boundary Cases, and Candidate Counts}
\label{app:normalization}

For any feature vector $v$, normalization uses
\begin{equation}
\hat v=\frac{v}{\max(\|v\|_2,\epsilon)}.
\label{eq:normalization}
\end{equation}

We use $\epsilon=10^{-6}$. For $T>1$, the first displacement is initialized as $\delta_1=(T-1)^{-1}\sum_{t=2}^T\delta_t$, and the standard deviation in Equation~(\ref{eq:frame-weights}) is the sample standard deviation. For $T=1$, we set $p_1=1$ and directly allocate the single-frame budget. The first frame's representative score uses cosine difference from its own mean, $c_{1,i}=1-\langle\hat x_{1,i},\hat\mu_1\rangle$.

For the first-stage budget, we floor each $B_1p_t$ and assign the remaining slots in descending order of fractional remainder, breaking ties by frame order. Applying the lower bound of one representative and upper bound $P$ gives the realized count $n_t$. We retain these counts after clipping, with $N_1=\sum_t n_t$. No redistribution follows this first-stage clipping, so active bounds can make $N_1$ differ from the nominal request $B_1$. Using realized counts keeps every downstream quota tied to available candidates.

\subsection{Cluster Assignment and Feature Alignment}
\label{app:alignment}

For a nonrepresentative token, the within-frame assignment is
\begin{equation}
a_t(i)=\arg\max_{j\in\mathcal R_t}
\langle\hat x_{t,i},\hat x_{t,j}\rangle,
\qquad i\notin\mathcal R_t.
\label{eq:assignment}
\end{equation}

Each representative belongs to its own cluster. Equation~(\ref{eq:aggregation}) averages the original features assigned to that cluster. Temporal comparisons determine the representative locations; aggregation combines features from the same frame. The assignment map is shared across the main visual embeddings and all DeepStack feature streams.

Candidates are assembled in frame order. Representative spatial-grid indices are used to construct multimodal positions, attaching each candidate to its representative location. Final selection applies a shared set of retained indices to the candidate embeddings, auxiliary features, and existing position encodings. The second stage performs direct pruning, so content aggregation occurs once during candidate construction.

\subsection{Probe Attention and Task Scores}
\label{app:attention}

The probe is a newline token inserted immediately after the final visual token. Let $A_{h,u,v}^{(\ell)}$ be the causal attention weight from query $u$ to visual candidate $v$ in head $h$ at scoring layer $\ell$. With probe position $u_0$, $H_a$ attention heads, and subsequent text-query positions $\mathcal U$, we compute
\begin{equation}
\tilde Q_v=\frac{1}{H_a}\sum_h A_{h,u_0,v}^{(\ell)},
\qquad
\tilde P_v=\frac{1}{H_a|\mathcal U|}
\sum_h\sum_{u\in\mathcal U}A_{h,u,v}^{(\ell)}.
\label{eq:attention}
\end{equation}

The two vectors are normalized over visual candidates before applying Equation~(\ref{eq:relevance}). The text suffix $\mathcal U$ includes the question, answer options, and template markers following the visual block. Causal masking gives the probe access to the preceding video and prompt while excluding the later question. The score retains its signed value for ranking.

The default reads attention from the third decoder block's input. A multi-layer extension averages the normalized reference distributions and task distributions separately across scoring layers, then applies the same correction formula. The probe and scoring formulation follow PriorTR~\citep{priortr}.

\subsection{Exact Integer Quota Allocation}
\label{app:quotas}

Given candidate counts $n_t$, final budget $K$, and minimum count $m$, initialize $\ell_t=\min(m,n_t)$. If $\sum_t\ell_t>K$, set all lower bounds to zero. The ideal quotas remain $q_t^*=Kn_t/N_1$, and the initial integer quotas are
\begin{equation}
q_t=\min\{n_t,\max(\ell_t,\lfloor q_t^*\rfloor)\}.
\label{eq:initial-quota}
\end{equation}

If $\sum_tq_t>K$, remove one slot from the eligible frame with the largest $q_t-q_t^*$, subject to $q_t>\ell_t$. If $\sum_tq_t<K$, add one slot to the eligible frame with the largest $q_t^*-q_t$, subject to $q_t<n_t$. Recompute deficits or surpluses after each adjustment and use frame order to break ties. Stop when the total equals $K$.

The procedure preserves each frame's bounds during every update. Because the effective lower bounds sum to at most $K$ and the capacities sum to $N_1\geq K$, an eligible adjustment exists whenever the total differs from $K$. Each update changes the total by one toward $K$, establishing termination with an exact final budget. With $m=1$ and sufficient budget, every nonempty processed temporal position retains a candidate.

\subsection{Optimal Selection for Fixed Scores and Quotas}
\label{app:optimality}

For candidate sets $\mathcal Y_t$, scores $s_v$, and assigned quotas $q_t$, the selection objective is
\begin{equation}
\max_{\mathcal S_1,\ldots,\mathcal S_T}
\sum_t\sum_{v\in\mathcal S_t}s_v,
\qquad
\mathcal S_t\subseteq\mathcal Y_t,
\quad |\mathcal S_t|=q_t.
\label{eq:selection-objective}
\end{equation}

The objective and constraints separate by frame. If a feasible selection contains a lower-scoring candidate while excluding a higher-scoring candidate from the same frame, exchanging them increases the objective and preserves the quota. Repeating these exchanges yields the $q_t$ highest-scoring candidates in every frame, proving Equation~(\ref{eq:selection}) optimal for this objective. Equal scores may admit multiple optimal selections.

\subsection{Invariance to Frame-Wise Score Calibration}
\label{app:score-invariance}

Fix the candidate sets and inherited quotas. For frame $t$, let $s'_v=a_t s_v+b_t$ with $a_t>0$. For any two candidates $u,v\in\mathcal Y_t$, the difference $s'_u-s'_v=a_t(s_u-s_v)$ has the same sign as $s_u-s_v$. Thus, the ordering within every frame, including ties, is unchanged. Applying the same tie-breaking rule returns the same $q_t$ candidates. Equivalently, for any feasible set in frame $t$, the transformed objective is $a_t\sum_{v\in\mathcal S_t}s_v+b_tq_t$; its maximizers are unchanged because $q_t$ is fixed. This property concerns the final selection operator with fixed candidates and quotas.

\section{Complete Procedure and Computational Accounting}
\label{app:procedure-cost}

\subsection{Compression and Inference Procedure}
\label{app:procedure}

\begin{algorithm}[t]
\caption{TeQCo: temporal quota-consistent video token compression.}
\label{alg:teqco}
\textbf{Input:} Visual features $X$, question text, retention ratios $\alpha$ and $\beta$, temperature $\tau$, and minimum frame count $m$.\par
\textbf{Output:} Answer generated from the compressed multimodal input.
\begin{enumerate}
    \item Pool and normalize frame features; compute first-order displacements.
    \item Standardize the displacements and allocate candidate budgets.
    \item Choose temporal representatives and assign other tokens within each frame.
    \item Aggregate each cluster, applying the same map to DeepStack features.
    \item Record realized candidate counts $n_t$ and their sum $N_1$.
    \item Insert the question-free probe and run shallow attention scoring.
    \item Compute $K$ and ideal quotas; allocate bounded integer quotas $q_t$.
    \item Select the highest-scoring $q_t$ candidates in each frame.
    \item Align visual embeddings, auxiliary features, and positions; retain all nonvisual text and remove the probe.
    \item Run a fresh prefill and generate the answer.
\end{enumerate}
\end{algorithm}

The scoring forward evaluates decoder blocks one through three and reads attention at the third block's input. Following selection, final prefill starts from the first decoder block with newly computed hidden states and KV caches. Visual encoding is completed before candidate construction.

\subsection{Cost Decomposition and Timing Boundary}
\label{app:cost}

Let $C_{\mathrm{LM}}(L,S)$ denote the cost of processing sequence length $S$ through $L$ language-model layers. TeQCo has approximate cost
\begin{equation}
\begin{aligned}
C_{\mathrm{TeQCo}}\approx{}&C_{\mathrm{ViT}}+C_{\mathrm{candidate}}
+C_{\mathrm{score}}(N_1+M+1)\\
&+C_{\mathrm{quota/select}}+C_{\mathrm{LM}}(L,K+M).
\end{aligned}
\label{eq:cost}
\end{equation}

The corresponding baseline cost is $C_{\mathrm{ViT}}+C_{\mathrm{LM}}(L,N+M)$. TeQCo shortens the sequence processed by the full decoder from $N+M$ to $K+M$. Net latency savings are the final-prefill savings after accounting for candidate construction, shallow scoring, and quota-based selection.

Within-frame matching has principal similarity cost proportional to $\sum_t(P-n_t)n_td$. Scoring extracts probe and text-query attention rows while its forward pass constructs a dense causal mask with quadratic sequence-length memory cost. Indexing, per-frame CPU/GPU transfers, and shallow decoder computation are included in inference accounting.

Model-inference time to first token (TTFT) includes full visual encoding, candidate construction, shallow scoring, quota selection, and final prefill. Video decoding and processor preprocessing precede this timed interval. This definition measures the combined compression and inference path.

\section{Configuration and Allocation Diagnostic}
\label{app:configuration}

\subsection{Default Configuration}
\label{app:defaults}
\label{sec:parameters}

\begin{table}[t]
\centering
\caption{Default TeQCo configuration.}
\label{tab:parameters}
\begin{tabularx}{\linewidth}{@{}lX@{}}
\toprule
Setting & Value \\
\midrule
Backbone & Frozen Qwen3-VL-8B-Instruct~\citep{qwen3vl} \\
Video sampling & Nominal 1 FPS, subject to duration and processor frame limits \\
Longest spatial edge & At most 512 pixels \\
Candidate retention $\alpha$ & 0.5 \\
Final retention $\beta$ & 0.5 of realized candidates \\
Content-change order & First order \\
Temperature $\tau$ & 8.0 on standardized changes \\
Minimum frame count $m$ & 1, enforced when feasible \\
Temporal reference & Same spatial location in the preceding frame \\
First-frame reference & Mean feature of the first frame \\
Candidate aggregation & Within-frame mean \\
Scoring location & Input to the third decoder block \\
Final selection & Top-$K$ within inherited frame quotas \\
\bottomrule
\end{tabularx}
\end{table}

The nominal retention product is $\alpha\beta=25\%$, targeting a reduction of approximately 75\% in the visual portion of the final-prefill input. Realized counts reflect first-stage clipping and integer allocation, and all nonvisual text remains in the sequence. Equation~(\ref{eq:cost}) accounts for the additional scoring computation.

\subsection{Deviation from the Inherited Allocation}
\label{app:diagnostic}

We measure the final temporal distribution using total-variation distance from the realized candidate proportions:
\begin{equation}
D_{\mathrm{budget}}=\frac{1}{2}\sum_t
\left|\frac{q_t}{K}-\frac{n_t}{N_1}\right|.
\label{eq:budget-drift}
\end{equation}

A value of zero indicates exact proportional inheritance. Positive values quantify the adjustment associated with integer counts and active coverage bounds. Together with the final frame counts and fraction of empty frames, this diagnostic makes the allocation relationship between compression stages directly measurable.

\end{document}